# Fast solving of Weighted Pairing Least-Squares systems

## Pierre Courrieu

*LPC, UMR 6146, CNRS-University of Provence, Marseille, France*

E-mail: Pierre.Courrieu@univ-provence.fr

**Abstract**

This paper presents a generalization of the "weighted least-squares" (WLS), named "weighted pairing least-squares" (WPLS), which uses a rectangular weight matrix and is suitable for data alignment problems. Two fast solving methods, suitable for solving full rank systems as well as rank deficient systems, are studied. Computational experiments clearly show that the best method, in terms of speed, accuracy, and numerical stability, is based on a special {1, 2, 3}-inverse, whose computation reduces to a very simple generalization of the usual "Cholesky factorization-backward substitution" method for solving linear systems.

**1. Introduction**

In this paper, we consider weighted least-squares problems in the following generalized form:

$$E(V;X,Y,W) = \sum_{i=1}^{m_1} \sum_{j=1}^{m_2} w_{ij} \left\| X_{i,:} V - Y_{j,:} \right\|^2,$$

$$C = \arg \min_{V \in R^{n_1 \times n_2}} E(V;X,Y,W), \tag{1}$$

where three matrices are given: $X \in R^{m_1 \times n_1}$, $Y \in R^{m_2 \times n_2}$, and $W = (w_{ij})$ is a rectangular $m_1 \times m_2$ "weighted pairing" matrix whose coefficients are non-negative real numbers. $X_{i,:}$ denotes the ith row of $X$, and $Y_{j,:}$ denotes the jth row of $Y$. Note that this generalization is not the same as the so-called "Generalized Least Squares" [10]. In the special case where $m_1 = m_2 = m$ and $W$ is a diagonal matrix, the above problem clearly reduces to an ordinary weighted least-squares problem, that is:

$$C = \arg\min_{V \in R^{n_1 \times n_2}} \sum_{i=1}^{m} w_{ii} \left\| X_{i,:} V - Y_{i,:} \right\|^2 = \arg\min_{V \in R^{n_1 \times n_2}} \left\| W^{1/2} XV - W^{1/2} Y \right\|^2. \quad (2)$$

In Section 2, we show that, in fact, every problem having the form (1) can be reduced to a problem having the form (2). In such problems, each equation of the least-squares system receives a specific weight that typically depends on some estimate of the reliability of the data used in that equation. The usual non-weighted case corresponds to $W = I$ (identity matrix). Ordinary weighted least-squares (2) are commonly used to solve regression problems with noisy data [12], and in "iteratively re-weighted least-squares" procedures for computing robust regression statistics such as M-estimators [2][7]. The generalization (1) is potentially relevant in "data alignment" problems, where there is no given one-to-one correspondence between $X$ data points and $Y$ data points (rows), but one has some non-negative "adequacy" or "plausibility" measure for each possible data pair, which is represented by $W$. Data alignment is a hard to solve problem commonly encountered in image processing and pattern recognition [6]. In this paper, ||.|| denotes the Euclidean norm for vectors, and the Frobenius norm for matrices. As in Matlab, the notation "a:b" denotes an index interval of bounds "a" and "b", and if the bounds are not specified (:), this corresponds to the whole index range.

Whenever $W^{1/2} X$ is of full column rank in (2), the solution to this problem is unique and the normal equations lead to the well-known result:

$$C = (X'WX)^{-1} X'WY, \tag{3}$$

where $X'$ denotes the transpose of $X$ (or the conjugate transpose in the complex case).

One of the fastest ways of numerically obtaining the factor $(X'WX)^{-1}$ that appears in (3) consists of computing a Cholesky factorization $LL'$ of the positive definite Gram matrix $X'WX$, then one inverts the upper triangular factor $L'$ by a simple backward substitution method, and one obtains $(X'WX)^{-1} = L'^{-1} L^{-1}$. However, if $W^{1/2}X$ is not of full column rank, then the above method does not work because the matrix $X'WX$ is singular, and in this case, the weighted least-squares system is said to be rank deficient. The solution of rank deficient systems requires more robust methods, which are also slower than the above mentioned, in general. Among the fastest methods, we can consider those based on the use of suitable generalized inverses, such as the popular Moore-Penrose inverse [1][4][8]. A solution to (2) is then:

$$C = (W^{1/2}X)^{\dagger} W^{1/2}Y = (X'WX)^{\dagger} X'WY, \tag{4}$$

where the Moore-Penrose inverse is known to provide the least-squares solution $C$ whose each column has the minimum Euclidean norm ([1], p. 109).

However, one must note that the solution of least-squares problems does not specifically require the use of the Moore-Penrose inverse, and that other types of generalized inverses, such as {1, 3}-inverses whose numerical computation is possibly faster, can as well be used. According to ([1], pp. 104-105), one has always a solution to (2) with:

$$C = (W^{1/2}X)^{(1,3)} W^{1/2}Y, \tag{5}$$

where $A^{(1,3)}$ denotes any {1, 3}-inverse of the matrix $A$ (see Section 3.2).

In fact, the problem of the computational cost is crucial in many practical applications, where one must repeatedly solve large least-squares systems. On the other hand, most practical problems lead to full rank systems that could be solved fast using (3), however, rank deficient systems can occasionally appear, and it is commonly not acceptable to obtain a "fatal error" diagnostic at run time. Thus, in order to optimize the performance of applications, we present in Section 3.2 a quite fast solution of type (4), and in Section 3.3 a solution of type (5) whose computational cost is similar to that of (3), which has the advantage of being fast while providing a suitable least-squares solution in all cases, even if the system is rank deficient. These solutions apply to problem (2) and to problem (1) as well.

## 2. The weighted pairing least-squares problem

In this section, we consider the generalization of the weighted least-squares (WLS) problem stated in (1), which we refer to as the "weighted pairing least-squares" (WPLS) problem.

**Theorem 1.** *Every WPLS problem of type (1) reduces to a WLS problem of type (2) since:*

$$\arg \min_{V \in R^{n1 \times n2}} \sum_{i=1}^{m1} \sum_{j=1}^{m2} w_{ij} \| X_{i,:} V - Y_{j,:} \|^2 = \arg \min_{V \in R^{n1 \times n2}} \sum_{i=1}^{m1} h_{ii} \| X_{i,:} V - Z_{i,:} \|^2,$$

*where:*

$H = (h_{ij})$ *is a diagonal matrix with diagonal coefficients* $h_{ii} = \sum_{j=1}^{m2} w_{ij}$, $1 \leq i \leq m1$,

$Z = H^\dagger W Y$,

$H^\dagger = (h_{ij}^\dagger)$ *is the Moore-Penrose inverse of* $H$, *with* $h_{ii}^\dagger = 1/h_{ii}$ *if* $h_{ii} > 0$, *and* $h_{ij}^\dagger = 0$ *if* $h_{ij} = 0$.

**Proof.**

Set $d_{ik} = \sum_{j=1}^{m2} w_{ij} y_{jk}^2 - h_{ii}^\dagger (\sum_{j=1}^{m2} w_{ij} y_{jk})^2$, $1 \leq i \leq m1$, $1 \leq k \leq n2$. (6)

Then one has:

$$\sum_{i=1}^{m1}(h_{ii}\|X_{i,:}V - Z_{i,:}\|^2 + \sum_{k=1}^{n2} d_{ik})$$

$$= \sum_{i=1}^{m1}\sum_{k=1}^{n2} h_{ii}(X_{i,:}V_{:,k} - h_{ii}^{\dagger}\sum_{j=1}^{m2} w_{ij} y_{jk})^2 + d_{ik}$$

$$= \sum_{i=1}^{m1}\sum_{k=1}^{n2}(h_{ii}^{1/2} X_{i,:}V_{:,k} - h_{ii}^{1/2} h_{ii}^{\dagger}\sum_{j=1}^{m2} w_{ij} y_{jk})^2 + d_{ik}$$

$$= \sum_{i=1}^{m1}\sum_{k=1}^{n2} h_{ii}(X_{i,:}V_{:,k})^2 - 2h_{ii}h_{ii}^{\dagger}(X_{i,:}V_{:,k})\sum_{j=1}^{m2} w_{ij} y_{jk} + h_{ii}h_{ii}^{\dagger 2}(\sum_{j=1}^{m2} w_{ij} y_{jk})^2 + d_{ik}$$

$$= \sum_{i=1}^{m1}\sum_{k=1}^{n2} h_{ii}(X_{i,:}V_{:,k})^2 - 2(X_{i,:}V_{:,k})\sum_{j=1}^{m2} w_{ij} y_{jk} + h_{ii}^{\dagger}(\sum_{j=1}^{m2} w_{ij} y_{jk})^2 + d_{ik}$$

$$= \sum_{i=1}^{m1}\sum_{k=1}^{n2}(\sum_{j=1}^{m2} w_{ij}(X_{i,:}V_{:,k})^2) - 2(\sum_{j=1}^{m2} w_{ij} y_{jk}(X_{i,:}V_{:,k})) + (\sum_{j=1}^{m2} w_{ij} y_{jk}^2)$$

$$= \sum_{i=1}^{m1}\sum_{k=1}^{n2}\sum_{j=1}^{m2} w_{ij}((X_{i,:}V_{:,k})^2 - 2(X_{i,:}V_{:,k})y_{jk} + y_{jk}^2)$$

$$= \sum_{i=1}^{m1}\sum_{j=1}^{m2} w_{ij}\|X_{i,:}V - Y_{j,:}\|^2.$$

Noting that the additional terms ($d_{ik}$) given by (6) are independent of $V$, one obtains Theorem 1. □

**Corollary 1.**

*(i) If $H^{1/2}X$ is of full column rank, then (1) has the unique solution:*

$$C = (X'HX)^{-1} X'HZ = (X'HX)^{-1} X'WY.$$

*(ii) No matter $H^{1/2}X$ is not of full column rank, (1) has the minimum norm solution:*

$$C = (H^{1/2}X)^{\dagger} H^{1/2}Z = (X'HX)^{\dagger} X'WY.$$

*(iii) No matter $H^{1/2}X$ is not of full column rank, (1) has all solutions of the form:*

$$C = (H^{1/2}X)^{(1,3)} H^{1/2}Z,$$

*where $H$ and $Z$ are defined as in Theorem 1.*

**Proof.** This directly follows from Theorem 1 and equation (3) for *(i)*, equation (4) for *(ii)*, and equation (5) for *(iii)*. □

## 3. Fast solutions based on generalized inverses

### 3.1 Generalized Cholesky factors

Several generalizations of the Cholesky factorization can be found in the literature. A well-known generalized Cholesky factorization for solving the so-called "augmented linear systems" is available in [9] and [11]. Another type of generalization of the Cholesky factorization has been proposed in [3], and this approach has been successfully used to define a fast numerical method for computing the Moore-Penrose inverse [4]. The fundamental result for the generalized Cholesky factorization is:

**Theorem 2** (from [3]). *Let $G$ be a symmetric positive semi-definite matrix of order $n \times n$. Then there is an upper triangular matrix $R$ such that $R'R = G$, $r_{ii} \geq 0$, $1 \leq i \leq n$, and if for an index $i$ one has $r_{ii} = 0$, then $r_{ij} = 0$, $1 \leq j \leq n$. Moreover, the matrix $R$ with these properties is unique.*

**Proof.** A proof of this is available in ([3], Theorem 4). □

The corresponding algorithm for computing the generalized Cholesky factor $R$ defined in Theorem 2 is a very simple variant of the usual Cholesky factorization algorithm, and its computational complexity is the same. However, the generalization has the advantage of providing a suitable factor in all cases, even if the matrix $G$ is singular.

**Algorithm 1.** *{Generalized Cholesky factor $R$ of the given matrix $G$}*

$r_{ij} \leftarrow 0, \quad 1 \leq i, j \leq n$             *{initialization of $R$}*

$r_{11} = \sqrt{g_{11}}$

for $j \leftarrow 2$ to $n$

    for $i \leftarrow 1$ to $j$

        if $i = j$ then $r_{ii} \leftarrow \sqrt{g_{ii} - \sum_{k=1}^{i-1} r_{ki}^2}$

        else if $r_{ii} > 0$ then $r_{ij} \leftarrow (g_{ij} - \sum_{k=1}^{i-1} r_{ki} r_{kj})/r_{ii}$

                          *{else $r_{ij} = 0$ as a result of the initialization}*.

By construction, the output of Algorithm 1 is an upper triangular factor $R$ with $r$ non-zero rows, and $n - r$ zero rows, where $r$ is the rank of $G$. The algorithm complexity is in $O(n^3)$, but the exact operations count depends on $r$ and the indices of zero rows. When $r = n$, this count is maximum, and it is equal to that of the classical Cholesky factorization (plus $n(n-1)/2$ low cost tests).

**3.2 Fast Moore-Penrose inverse based solution**

Using Algorithm 1, one can define a fast method for computing the Moore-Penrose inverse of every finite matrix. Before examining this method, we rapidly recall some definitions and notations concerning generalized inverses.

Every finite matrix $A$ has (possibly an infinite number of) generalized inverses (hereafter denoted $B$) that satisfy one or several of the following four Penrose equations:

$$ABA = A \qquad (P1)$$

$$BAB = B \qquad (P2)$$

$$(AB)' = AB \qquad (P3)$$

$$(BA)' = BA \qquad (P4)$$

Every matrix $B$ that satisfies the equation set $\{Pi, Pj,...\}$ is said to be a $\{i, j,...\}$-inverse of $A$, and it is usually denoted $A^{(i,j,...)}$. The Moore-Penrose inverse of $A$ is the unique matrix $A^\dagger = A^{(1,2,3,4)}$. For a complete explanation, the reader can see [1].

There are several methods for computing the Moore-Penrose inverse, the most usual being based on the singular values decomposition (SVD). This method is numerically very stable, however it is computationally heavy and hardly usable in many practical applications. Another usual method is based on Gram-Schmidt orthonormalization, which is clearly faster than SVD. However, the classical Gram-Schmidt orthonormalization (CGS or GSO) is known to be numerically instable. A simple and effective remediation to this drawback has been proposed in the form of a re-orthogonalization additional step, leading to the CGS2 method [5]. However, the additional step in CGS2 slows down the process, while it has been observed that CGS is not the fastest method for computing the Moore-Penrose inverse [4]. In fact, it turned out that among the most usual methods, including Greville's method, SVD, CGS/GSO, and iterative methods, the fastest known numerical method for computing the Moore-Penrose inverse is based on Algorithm 1 and on the following result [4]:

**Theorem 3** (from [4]). *Let $A$ be a $m \times n$ matrix, with $m \geq n$, set $G = A'A$, compute the generalized Cholesky factorization $G = R'R$ using Algorithm 1, remove all zero rows from $R$,* *which results in a full row rank matrix $S$ of size $r \times n$, with $r \leq n$, and such that $S'S = G$. Then:*

$$A^\dagger = S'(SS')^{-1}(SS')^{-1}SA'.$$

**Proof.** The proof is available in [4]. Since it is short, we provide it hereafter.

We start with equation (3.2) from [8], that is:

$$(EF)^\dagger = F'(E'EFF')^\dagger E'. \qquad (7)$$

Setting $E = A$, and $F = I$ in (7), one obtains:

$$A^\dagger = (A'A)^\dagger A' = G^\dagger A'.$$

Setting $E = S'$, and $F = S$ in (7), one obtains:

$$G^\dagger = (S'S)^\dagger = S'(SS'SS')^\dagger S = S'(SS')^{-1}(SS')^{-1}S, \qquad (8)$$

since $SS'$ is invertible because $S$ is of full row rank. □

If $A$ is a $m \times n$ matrix, with $m < n$, it suffices to use the relation $A^\dagger = ((A')^\dagger)'$. Note also that (8) provides a simple formula for the Moore-Penrose inverse of any symmetric positive semi-definite matrix, and that if $S$ is of full rank $r = n$, then $G^\dagger = G^{-1}$.

**Corollary 2.** *Set $A = H^{1/2}X$ in Theorem 3, then the minimum norm solution of (1) is:*

$$C = S'(SS')^{-1}(SS')^{-1}SX'WY,$$

*where $S$ is defined as in Theorem 3.*

**Proof.** This immediately follows from Theorem 3 and Corollary 1 *(ii)*. □

### 3.3 Fast {1, 2, 3}-inverse based solution

Although Corollary 2 provides a fast solution to (1), this is not necessarily the fastest way of solving this problem. Moreover, observing equation (8), one can suspect a potential numerical instability whenever the matrix $SS'$ is ill-conditioned, worsened by the fact that the factor $(SS')^{-1}$ is repeated. In this section, we describe a {1, 2, 3}-inverse based solution to (1) whose computational complexity is similar to that of (3), using Algorithm 1 and a simple variant of the backward substitution method for inverting upper triangular matrices. The simplicity of this solution allows us to expect not only faster computation, but also better numerical stability than the Moore-Penrose inverse based approach. We first define the generalization of the backward substitution method for computing generalized inverses of generalized Cholesky factors as they are defined in Theorem 2. The computational complexity of the generalized algorithm is the same as that of the original backward substitution method ($O(n^3)$). The algorithm is designed to solve in $U$ the following equation:

$$RU = I_R, \qquad (9)$$

where $I_R$ is a diagonal $n \times n$ matrix whose ith diagonal coefficient is equal to 0 if the ith row of $R$ is zero, else this diagonal coefficient is equal to 1. The algorithm to solve (9) is then:

**Algorithm 2.** *{{1, 2, 3}-inverse $U$ of the given generalized Cholesky factor $R$}*

$u_{ij} \leftarrow 0, \quad 1 \le i, j \le n$              *{initialization of $U$}*

for $j \leftarrow n$ downto 1

     if $r_{jj} \ne 0$ then           *{note: this test is optional}*

         for $i \leftarrow j$ downto 1

             if $r_{ii} \ne 0$ then

                 if $i = j$ then $u_{ii} \leftarrow 1/r_{ii}$

                 else $u_{ij} \leftarrow -(\sum_{k=i+1}^{j} r_{ik} u_{kj})/r_{ii}$.

The test at the third line of Algorithm 2 is optional because it has no influence on the result. However, including this test allows to save a number of useless floating-point operations (whose result is zero) whenever $R$ is singular.

By construction, the output of Algorithm 2 is an upper triangular matrix $U$ that has the property that if the ith row of $R$ is zero, then both the ith row and the ith column of $U$ are zero. Note that the rank of $U$ is equal to the rank of $R$, which is itself equal to the rank of the factorized matrix $G$ and to the trace of $I_R$. Note also that if $R$ is not invertible (in the usual sense), then $UR \neq I_R$, however, $UR$ is always idempotent since $URUR = UI_R R = UR$.

**Theorem 4.** *Let $R$ be a generalized Cholesky factor as defined in Theorem 2, let $U$ be the corresponding output of Algorithm 2, then $U$ is a {1, 2, 3}-inverse of $R$.*

**Proof.**

$U$ is a {1}-inverse of $R$ since $RUR = I_R R = R$,

$U$ is a {3}-inverse of $R$ since $RU = (RU)' = I_R$,

$U$ is a {2}-inverse of $R$ since $U$ is a {1}-inverse of $R$ and has the same rank as $R$, then the conclusion follows from ([1], p. 46). Alternatively, one can easily verify that $URU = UI_R = U$. □

**Theorem 5.** *Let $A$ be a $m \times n$ matrix, with $m \geq n$, set $G = A'A$, compute the generalized Cholesky factorization $G = R'R$ using Algorithm 1, compute $U = R^{(1,2,3)}$ using Algorithm 2. Then:*

*(i) The equation $A = QR$ has a solution in $Q$ such that $Q'Q = I_R$. This solution is $Q = AU$.*

*(ii) The matrix $B = UU'A'$ is a {1, 2, 3}-inverse of $A$.*

**Proof.**

$$Q'Q = U'A'AU = U'R'RU = (RU)'RU = I_R I_R = I_R,$$

$$AU = QRU = QI_R = Q,$$

which proves *(i)*.

Since *(i)* implies that $B = UQ'$, one has:

$B$ is a {1}-inverse of $A$ since $ABA = QRUQ'QR = QI_R I_R R = QR = A$,

$B$ is a {2}-inverse of $A$ since $BAB = UQ'QRUQ' = UI_R I_R Q' = UQ' = B$,

$B$ is a {3}-inverse of $A$ since $AB = QRUQ' = QI_R Q' = (AB)'$,

which proves *(ii)*, and then completes the proof of Theorem 5. $\square$

**Corollary 3.** *Set $A = H^{1/2}X$ in Theorem 5, then a solution to (1) is given by:*

$$C = UU'X'HZ = UU'X'WY,$$

*where $U$ is defined as in Theorem 5.*

**Proof.** This immediately follows from Theorem 5 *(ii)* and Corollary 1 *(iii)*. $\square$

One can note that if $H^{1/2}X$ is of full column rank, then $U = R^{-1}$, and the solution provided by Corollary 3 is equal to that of Corollary 1 *(i)*. Moreover, if $H^{1/2}X$ is of column rank $r \leq n1$, then each column of the solution $C$ to problem (1) provided by Corollary 3 has at most $r$ non-zero coefficients, since the first factor ($U$) of the solution has $n1 - r$ zero rows. Note, however, that one can find particular examples showing that the above solution is not always the one having the minimum number of non-zero coefficients.

## 4. Computational test

### 4.1 Implementation of methods

The methods defined in Corollary 2 and Corollary 3 for solving (1) have been implemented in Matlab code (version 7.5), and are listed in Appendix 1. The Matlab function corresponding to Corollary 2 is named "WPLSdagger", and the Matlab function corresponding to Corollary 3 is named "WPLS123". This makes available various implementation details that are not specified in the formal definition of algorithms, such as the way of testing the equivalence to zero of floating point diagonal coefficients, or the way of avoiding an a posteriori removing of zero rows in Corollary 2 solution. In order to make the performance of the two tested methods comparable, we avoided the use of high level Matlab operators such as "inv()", "chol()", or "\", whose implementation is hidden and compiled.

### 4.2 Test problems

In order to test the performance of methods for solving (1) in terms of speed, accuracy, and numerical stability, we must build test problems in a way that allows strict control of relevant characteristics such as the size and the rank of the equation system, the exact weighted least-squares residue norm, and the ratio of extreme non-zero singular values of the system matrix (which can be seen as a kind of generalized condition number). Note that the solution ($C$) itself is not relevant for comparisons, since it is not unique in the case of rank deficient systems. Building coherent test problems having all required properties is not so easy, and we propose the following method.

First, one chooses the size parameters $m1$, $n1$, $m2$, $n2$, the rank parameter $r \leq n1$, and the maximum ratio, denoted $\kappa_r$, of non-zero eigenvalues of the Gram matrix $X'HX$ to be built. For practical reasons, one must choose the size parameters such that $m2 \geq m1 > n1$. Then one builds two orthogonal Householder matrices:

$$M = I - 2\frac{uu'}{u'u}, \text{ with a random column vector } u \in R^{m1},$$

$$N = I - 2\frac{vv'}{v'v}, \text{ with a random column vector } v \in R^{n1},$$

where the identity matrices ($I$) have the appropriate size in each case. One also builds a $r \times r$ diagonal matrix $D$, whose ith diagonal coefficient is equal to $\kappa_r^{(r-i)/2(r-1)}$, $1 \leq i \leq r$. Then one selects the first $r$ columns of $M$ and the first $r$ rows of $N$, and one builds the matrix:

$$A = M_{:,1:r} D N_{1:r,:}.$$

The $m1 \times n1$ matrix $A$ is of rank $r$, the greatest eigenvalue of $A'A$ is equal to $\kappa_r$, and the lowest non-zero eigenvalue of $A'A$ is equal to 1. Thus, we can set $X'HX = A'A$, that is $H^{1/2}X = A$.

For the next step, one builds a random $(m1 - r) \times n2$ real matrix $F$, and one sets:

$$P = M_{:,(r+1):m1} F,$$

where we note that the columns of $P$ are orthogonal to those of $A$.

One can now build a suitable diagonal matrix $H = (h_{ii}), 1 \leq i \leq m1$, by taking:

$$h_{ii} = \max\left(|\sum_{j=1}^{n1} a_{ij}|, |\sum_{j=1}^{n2} p_{ij}|\right)^2,$$

which guarantees that both $A$ and $P$ can be factorized with $H^{1/2}$ as the first factor, in order to build a coherent problem, and one obtains the first matrix of problem (1) by:

$$X = (H^{1/2})^\dagger A.$$

For the next step, one builds a random $n1 \times n2$ real matrix $V$, and one sets:

$$HZ = H^{1/2}(AV + P).$$

It remains to build a $m1 \times m2$ matrix $W$, with non-negative coefficients, such that $\sum_{j=1}^{m2} w_{ij} = h_{ii}, 1 \leq i \leq m1$, and such that there is a $m2 \times n2$ matrix $Y$ that is solution of the equation $WY = HZ$. Unfortunately, there is no available deterministic method for factorizing $HZ$ in a suitable way, thus we must use a random "trials and errors" approach, as follows. Repeat the following four steps until $WY = HZ$ (at the working precision):

- build a $m1 \times m2$ matrix $T = (t_{ij})$ with non-negative random coefficients,

- compute the diagonal matrix $K$ with $k_{ii} = \sum_{j=1}^{m2} t_{ij}, 1 \leq i \leq m1$,

- set $W = HK^{-1}T$,

- set $Y = W^{\dagger}HZ$.

The Moore-Penrose inverse $W^{\dagger}$ can be obtained using an accurate (slow) SVD method. In general, one obtains a suitable solution quite rapidly when $m2 > m1$, and $\kappa_r$ is not too large. However, one can observe that the above process frequently fails for large systems if $\kappa_r > 2^{12}$, which seems to be a practical limit for generating problems in common computational environments such as Matlab.

We have now suitable matrices $X$, $Y$, and $W$ for problem (1), and it remains to compute the exact weighted sum of quadratic residues (i.e. the minimized $E$ function of (1)) corresponding to this problem as a reference value for testing the accuracy of solving methods. In order to do this, we use the fact that the columns of the matrix $P$ are orthogonal to those of $A$, and the proof of Theorem 1. Then one obtains:

$$E_{exact} = \|P\|^2 + \sum_{i=1}^{m1} \sum_{k=1}^{n2} d_{ik},$$

where the additional terms ($d_{ik}$) are defined as in (6).

**4.3 Results**

Using the procedure described in Section 4.2, we generated various problems of type (1) with the parameter sets $n_1 = \{128, 256, 512\}$, $\kappa_r = \{16, 256, 4096\}$, $r = \{n_1, \frac{7}{8}n_1\}$ (corresponding to "full rank" and "rank deficient" systems, respectively), while $m_1 = 2n_1$, $m_2 = 2m_1$, $n_2 = 32$. Using all parameter combinations, one obtained 18 types of problems, and 10 problems of each type were randomly generated. Each problem was solved by both the WPLSdagger function (fast Moore-Penrose inverse based solution), and the WPLS123 function (fast {1, 2, 3}-inverse based solution). In each case, the solving time was recorded in milliseconds (in Matlab 7.5, on a MacBook computer), and the accuracy of each method was measured by $(E_{method} - E_{exact})/E_{exact}$. The mean solving times are reported in Table 1, and the mean accuracy values are reported in Table 2. All differences between the two methods are statistically significant ($p<.01$) using the sign test.

**Table 1.** Mean solving time (in milliseconds) of WPLS problems by the two methods.

| $n_1$ : | 128 | | | 256 | | | 512 | | |
| $\kappa_r$ : | 16 | 256 | 4096 | 16 | 256 | 4096 | 16 | 256 | 4096 |
|---|---|---|---|---|---|---|---|---|---|
| Full rank | | | | | | | | | |
| WPLSdagger | 126 | 126 | 127 | 608 | 610 | 607 | 3438 | 3438 | 3434 |
| WPLS123 | 109 | 109 | 109 | 523 | 523 | 523 | 2868 | 2869 | 2869 |
| Rank deficient | | | | | | | | | |
| WPLSdagger | 103 | 102 | 102 | 502 | 497 | 498 | 2805 | 2803 | 2803 |
| WPLS123 | 89 | 89 | 89 | 433 | 434 | 433 | 2412 | 2412 | 2468 |

As one can see in Table 1, the {1, 2, 3}-inverse based method is faster than the fast Moore-Penrose inverse based method, in all cases. One can also note that rank deficient systems are solved faster than full rank systems of the same size by both methods, which is a consequence of the zeroing of certain rows in Algorithm 1. Moreover, an inspection of Table 2 clearly shows that the {1, 2, 3}-inverse based method is accurate in all cases, while the fast

Moore-Penrose inverse based method is less accurate and highly sensitive to the parameter $\kappa_r$, thus numerically instable. In summary, it seems that the {1, 2, 3}-inverse based method is globally preferable to other known methods suitable for solving problem (1), except if, for some particular reason, one requires a minimum norm solution. However, in this last case, it is certainly preferable to use an accurate and numerically stable method for computing the Moore-Penrose inverse, but the price to be paid for this is, in general, a quite long computation time.

**Table 2.** Mean accuracy of the two methods in solving WPLS problems.

| $n1$ : | | 128 | | | 256 | | | 512 | |
| $\kappa_r$ : | 16 | 256 | 4096 | 16 | 256 | 4096 | 16 | 256 | 4096 |
|---|---|---|---|---|---|---|---|---|---|
| Full rank | | | | | | | | | |
| WPLSdagger | 8.9E-6 | 9.9E-3 | 1.532 | 1.4E-6 | 1.3E-3 | 0.436 | 0.2E-6 | 0.2E-3 | 0.099 |
| WPLS123 | $< 10^{-12}$ | $< 10^{-12}$ | $< 10^{-12}$ | $< 10^{-12}$ | $< 10^{-12}$ | $< 10^{-12}$ | $< 10^{-12}$ | $< 10^{-12}$ | $< 10^{-12}$ |
| Rank deficient | | | | | | | | | |
| WPLSdagger | 3.6E-6 | 11.9E-3 | 2.872 | 0.9E-6 | 2.1E-3 | 0.570 | 0.1E-6 | 0.3E-3 | 0.137 |
| WPLS123 | $< 10^{-12}$ | $< 10^{-12}$ | $< 10^{-12}$ | $< 10^{-12}$ | $< 10^{-12}$ | $< 10^{-12}$ | $< 10^{-12}$ | $< 10^{-12}$ | $< 10^{-12}$ |

## 5. Conclusion

We have first generalized "weighted least-squares" (WLS) to "weighted pairing least-squares" (WPLS) problems. This generalization, which allows using a rectangular weight matrix, includes, as particular cases, the classical weighted and non-weighted least-squares problems, and it is more particularly suitable in the framework of data alignment problems. We have shown that WPLS problems can always be reduced to problems having the same form as WLS problems, and we have studied two fast methods for solving such problems in the case of rank deficient systems as well as of full rank systems. Numerical experiments clearly showed that the best solving method, in terms of speed, accuracy, and numerical stability, is based on a special {1, 2, 3}-inverse whose computation is very simple. In contrast, approaches based on the Moore-Penrose inverse lead to slow computation, or alternatively to numerical instability.

## Appendix 1

The following codes are provided for example, and for academic use only. The code is not optimized and exception cases are not managed.

*Matlab code (version 7.5) of the WPLSdagger function:*

```
function [C,Emethod,time] = WPLSdagger(X,Y,W)
% Moore-Penrose inverse based solution of a WPLS problem
tic    % start the clock
[m1,n1]=size(X); [m2,n2]=size(Y); H=sum(W,2); g=X'*((H*ones(1,n1)).*X);
% s = full row rank generalized Cholesky factor of g
tol=n1*eps(norm(g,inf));
s=zeros(n1,n1); ii=0;
for i=1:n1
    ii=ii+1;
    v=g(i,i:n1)-s(1:(ii-1),i)'*s(1:(ii-1),i:n1);
    if v(1)>tol
        s(ii,i)=sqrt(v(1));
        if i<n1
            s(ii,(i+1):n1)=v(2:end)/s(ii,i);
        end
    else ii=ii-1; end
end
rs=ii; s=s(1:rs,:);
% r = classical upper Cholesky factor of ss'
g=s*s';
r=zeros(rs,rs);
for i=1:rs
    v=g(i,i:rs)-r(1:(i-1),i)'*r(1:(i-1),i:rs);
    r(i,i)=sqrt(v(1));
    if i<rs
        r(i,(i+1):rs)=v(2:end)/r(i,i);
    end
end
% u = classical inverse of r
u=zeros(rs,rs);
for j=rs:-1:1
        for i=j:-1:1
                if i==j
                    u(i,i)=1/r(i,i);
                else
                    u(i,j)=-r(i,(i+1):j)*u((i+1):j,j)/r(i,i);
                end
        end
end
% iss = inverse of ss'
iss=u'*u;
% solution
C=s'*iss*iss*s*X'*W*Y;
time=toc;    % record the solving time
% compute the weighted sum of quadratic residues
```

```
XC=X*C;
Emethod=0;
for i=1:m1
    for j=1:m2
        Emethod=Emethod+W(i,j)*sum((XC(i,:)-Y(j,:)).^2,2);
    end
end
end
```

*Matlab code (version 7.5) of the WPLS123 function:*

```
function [C,Emethod,time] = WPLS123(X,Y,W)
% {1,2,3}-inverse based solution of a WPLS problem
tic     % start the clock
[m1,n1]=size(X); [m2,n2]=size(Y); H=sum(W,2); g=X'*((H*ones(1,n1)).*X);
% r = generalized Cholesky factor of g
tol=n1*eps(norm(g,inf));
r=zeros(n1,n1);
for i=1:n1
    v=g(i,i:n1)-r(1:(i-1),i)'*r(1:(i-1),i:n1);
    if v(1)>tol
        r(i,i)=sqrt(v(1));
        if i<n1
            r(i,(i+1):n1)=v(2:end)/r(i,i);
        end
    end
end
% u = {1,2,3}-inverse of r
u=zeros(n1,n1);
for j=n1:-1:1
    if r(j,j)~=0
        for i=j:-1:1
            if r(i,i)~=0
                if i==j
                    u(i,i)=1/r(i,i);
                else
                    u(i,j)=-r(i,(i+1):j)*u((i+1):j,j)/r(i,i);
                end
            end
        end
    end
end
% solution
C=u*u'*X'*W*Y;
time=toc;     % record the solving time
% compute the weighted sum of quadratic residues
XC=X*C;
Emethod=0;
for i=1:m1
    for j=1:m2
        Emethod=Emethod+W(i,j)*sum((XC(i,:)-Y(j,:)).^2,2);
    end
end
end
```

**References**

[1] A. Ben-Israel and T.N.E. Greville, *Generalized Inverses: Theory and Applications (2nd Ed.)*, Springer-Verlag, New York, 2003.

[2] K.P. Bube and R.T. Langan, Hybrid $l^1/l^2$ minimization with applications to tomography, *Geophysics* **62(4)** (1997) 1183-1195.

[3] P. Courrieu, Straight monotonic embedding of data sets in Euclidean spaces, *Neural Networks* **15** (2002) 1185-1196.

[4] P. Courrieu, Fast computation of Moore-Penrose inverse matrices, *Neural Information Processing – Letters and Reviews* **8(2)** (2005) 25-29.

[5] L. Giraud, J. Langou, M. Rozloznik, and J. van den Eshof, Rounding error analysis of the classical Gram-Schmidt orthogonalization process, *Numerische Mathematik* **101** (2005) 87-100.

[6] S.-H. Lai, Robust image matching under partial occlusion and spatially varying illumination change, *Computer Vision and Image Understanding* **78** (2000) 84-98

[7] R.A. Maronna, R.D. Martin, and V.J. Yohai, *Robust Statistics: Theory and Methods*, John Wiley & Sons Ltd, New York, 2006.

[8] M.A. Rakha, On the Moore-Penrose generalized inverse matrix, *Applied Mathematics and Computation* **158** (2004) 185-200.

[9] W. Wang and J. Zhao, Perturbation analysis for the generalized Cholesky factorization, *Applied Mathematics and Computation* **147** (2004) 601-606.

[10] J.Y. Yuan, Numerical methods for generalized least squares problems, *Journal of Computational and Applied Mathematics* **66** (1996) 571-584.

[11] J. Zhao, The generalized Cholesky factorization method for saddle point problems, *Applied Mathematics and Computation* **92** (1998) 49-58.

[12] T. Zhou and D. Han, A weighted least squares method for scattered data fitting, *Journal of Computational and Applied Mathematics* **217** (2008) 56-63.